\documentclass[twocolumn,showpacs,amsmath,amssymb,prd]{revtex4}
          
\usepackage{graphicx}
\usepackage{dcolumn}
\usepackage{bm}

\def\beq{\begin{equation}}
\def\enq{\end{equation}}
\def\ba{\begin{eqnarray}}
\def\ea{\end{eqnarray}}
\def\<{\langle}
\def\>{\rangle}

\newcommand{\ev}{\hbox{ eV}}
\def\aprle{\buildrel < \over {_{\sim}}}

\begin{document}
\input{epsf}

\title{Oscillation effects on high-energy neutrino fluxes from 
astrophysical hidden sources}

\author{Olga Mena$^1$, Irina Mocioiu$^2$ and Soebur Razzaque$^{3,2}$}
\affiliation{$^1$ INFN Sez.\ di Roma,                     
Dipartimento di Fisica, Universit\`{a} di Roma``La Sapienza'', P.le
A.~Moro, 5, I-00185 Roma,Italy}
\affiliation{$^2$Department of Physics, 
Pennsylvania State University, University Park, PA 16802, USA}
\affiliation{$^3$Department of Astronomy \& Astrophysics, 
Pennsylvania State University, University Park, PA 16802, USA}

\begin{abstract}
High-energy neutrinos are expected to be produced in a vareity of
astrophysical sources as well as in optically thick hidden sources. We
explore the matter-induced oscillation effects on emitted neutrino
fluxes of three different flavors from the latter class. We use the
ratio of electron and tau induced showers to muon tracks, in upcoming
neutrino telescopes, as the principal observable in our analysis. This
ratio depends on the neutrino energy, density profile of the sources
and on the oscillation parameters. The largely unknown flux
normalization drops out of our calculation and only affects the
statistics. For the current knowledge of the oscillation parameters we
find that the matter-induced effects are non-negligible and the
enhancement of the ratio from its vacuum value takes place in an
energy range where the neutrino telescopes are the most
sensitive. Quantifying the effect would be useful to learn about the
astrophysics of the sources as well as the oscillation parameters. If
the neutrino telescopes mostly detect diffuse neutrinos without
identifying their sources, then any deviation of the measured flux
ratios from the vacuum expectation values would be most naturally
explained by a large population of hidden sources for which
matter-induced neutrino oscillation effects are important.
\end{abstract}


\date{\today}
\maketitle

\section{Introduction}

High-energy ($\gtrsim 100$ GeV) neutrino emission has been predicted
from several types of astrophysical sources such as active galactic
nuclei (AGNs), gamma-ray bursts (GRBs), core-collapse supernovae
(SNe), supernova remnants (SNRs), microquasars,
etc. \cite{Dermer:2006xt}. All these sources have been observed in
electromagnetic wave bands ranging from radio to high-energy
$\gamma$-rays. Low energy ($\sim 10$ MeV) thermal neutrinos have been
detected from a nearby core-collapse SN 1987A
\cite{Hirata:1987hu}. High-energy neutrinos may be detected from a
much longer distance because of an increasing interaction
cross-section with energy
\cite{Frichter:1994mx,gqrs,Gandhi:1998ri}. Upcoming kilometer scale
ice/water Cherenkov detectors such as IceCube \cite{Achterberg:2006md}
in Antarctica and its counterpart in the Mediterranean, called KM3NeT
\cite{Katz:2006wv}, will thus open up a new observation window in
high-energy neutrinos.

Neutrinos, unlike their electromagnetic counterparts, may carry
information on temperature, density, etc. from deep inside the
astrophysical sources. Their detection may also be used to probe
neutrino flavor oscillation parameters in matter and in vacuum as they
propagate inside the sources and over astrophysical distances to reach
Earth. The results will be complementary to accelerator and reactor
based experiments. Particular examples have been carried out in detail
for $\sim 10$ MeV thermal neutrinos from core-collapse SNe
\cite{Dighe:1999bi}.

Oscillations of high energy neutrinos have also been extensively
discussed in Refs.\
\cite{Rodejohann:2006qq,Winter:2006ce,Serpico:2005sz,serpico,Costantini:2004ap,Xing:2006uk,Cirelli:2005gh,Fogli:2006jk,lunardinismirnov}.
The sensitivity of high energy neutrinos to oscillation parameters has
been explored in Refs.\
\cite{Rodejohann:2006qq,Winter:2006ce,Serpico:2005sz,serpico,Costantini:2004ap,Xing:2006uk}
Matter effects in the context of neutrinos from dark matter
annihilation have been considered in Ref.\
\cite{Cirelli:2005gh}. Oscillations of solar atmosphere neutrinos have
been discussed in Ref.\ \cite{Fogli:2006jk}. Matter effects for high
energy neutrinos are generally small. In order to get significant
effects it is necessary to encounter a resonant density, as well as to
go through a minimum matter width, as shown in Ref.\
\cite{lunardinismirnov}. These conditions are usually not satisfied
for optically thin sources.

In this paper we explore flavor oscillation effects on high-energy
neutrinos from optically thick sources where matter effects can be
larger. These neutrinos are produced deep inside the astrophysical
sources (so-called ``hidden sources'') and as such their emission from
the source is not accompanied by any high-energy electromagnetic
component. Examples of hidden sources may be highly relativistic GRB
jets as they are formed inside collapsed stars and start burrowing
their way through the stellar envelope
\cite{Meszaros:2001ms,Razzaque:2003uv}; semi-relativistic jets inside
core-collapse SNe which may impact and disrupt the envelope but choke
inside \cite{Razzaque:2004yv}; and core dominated AGNs
\cite{Alvarez-Muniz:2004uz,Stecker:2005hn}. The first two cases are
motivated by the observation of a number of GRBs associated with
supernovae \footnote{confirmed detections are: GRB 980425/SN 1998bw,
GRB 021211/SN 2002lt GRB 030329/SN 2003dh, GRB 0131203/SN 2003lw and
GRB 060218/SN 2006aj. Many other GRBs show a SN bump in their
afterglow light curve but no spectroscopic data.} which supports the
collapsar model of GRBs \cite{MWH01}. The collapsar model ties the
GRBs and supernovae in a common thread, both originating from
core-collapse of massive stars. While observed GRBs are endowed with
highly relativistic jets, many more such collapses are expected to
produce mildly relativistic jets. In both cases the jet is launched
from inside the star \cite{MWH01, MR01, WM03} and burrows through the
stellar interior while the envelope is still intact.  High energy
neutrinos, produced by collisions of plasma materials (so called
``internal shocks'') in the jet, are emitted following a density
gradient while the jet itself may or may not break through the
envelope. High-energy neutrinos are produced via decays of pions and
kaons created by hadronic ($pp$) and/or photo-hadronic ($p\gamma$)
interactions of shock-accelerated protons in these examples. We
concentrate on matter enhancement of the electron (anti)neutrino flux
due to the small mixing angle $\theta_{13}$ as these neutrinos
propagate to the stellar surface from their production site. Matter
effects have not been previously discussed for these sources and they
provide an opportunity for identifying and studying such
``hidden''sources through the neutrinos they emit. We show in
particular the energy range where this effect may be detectable by
neutrino telescopes, and we also comment on possible extraction of
neutrino properties such as mass hierarchy and the CP violation phase.

The organization of the paper is as follows: We outline the
astrophysical source model and neutrino flux parametrization in
Sec.\ \ref{sec.astro}. We discuss neutrino oscillations, our numerical
approach to calculate the effects and analytic expectations in
Sec.\ \ref{sec.oscillation}. Detection and our results are given in
Sec.\ \ref{sec.detection} and in Sec.\ \ref{sec.analysis}
respectively. Conclusions and outlooks are given in
Sec.\ \ref{sec.conclusion}.

\section{Astrophysics of the source}
\label{sec.astro}

The particular hidden source model we employ for calculation purpose
is a jetted core collapse supernova model in Ref.\
\cite{Razzaque:2004yv}.  Our treatment may however be applied to any
generic case of a neutrino flux and matter density profile. The
presupernova star is a blue supergiant (BSG) with a radiative hydrogen
envelope which is capable to produce a Type II or Ib SN. The hydrogen
envelope sits on a helium core of radius $r_{\rm He} \sim 10^{11}$
cm. In case of significant stellar mass loss, the presupernova stars
may have no hydrogen envelope left at all and produce Type Ic SNe, a
fraction of which are now strongly believed to produce GRBs as
well. However, we do not investigate this case as the relativistic jet
front ($r_{\rm jet} \sim 10^{10.8}$ cm) is too close to the stellar
surface to have any significant amount of material in front
\cite{Razzaque:2004yv}.

An analytic form of the density distribution near the edge of each
layer of a star with polytropic structure is given by $\rho = \rho_1
(R_\star/r -1)^n$ \cite{matzmckee99}. Here $R_\star$ is the star's
radius and the polytropic index $n=3$ for a radiative envelope with
constant Thomson opacity. For a helium star of $R_\star \approx
10^{11}$~cm, $\rho_1 \approx 2$~g~cm$^{-3}$ and $n=3$ were found by
fitting data from SN 1998bw \cite{wooslangweav93}. The presupernova
star of SN 1987A is a BSG with $R_\star \approx 3\cdot 10^{12}$~cm,
$\rho_1 \approx 3\cdot 10^{-5}$~g~cm$^{-3}$ and $n=3$ \cite{shignom90,
arnett91}. Below we write 3 models of the density profile that we use
for a BSG of $R_\star=3\cdot 10^{12}$~cm. All models are normalized to
give the same density $\rho = 2(10^{11}/10^{10.8}-1)^3
=0.4$~g~cm$^{-3}$, as in the case of a helium star, at $r=r_{\rm jet}
= 10^{10.8}$~cm.

\begin{figure} [ht]
\centerline{\epsfxsize=3.4in \epsfbox{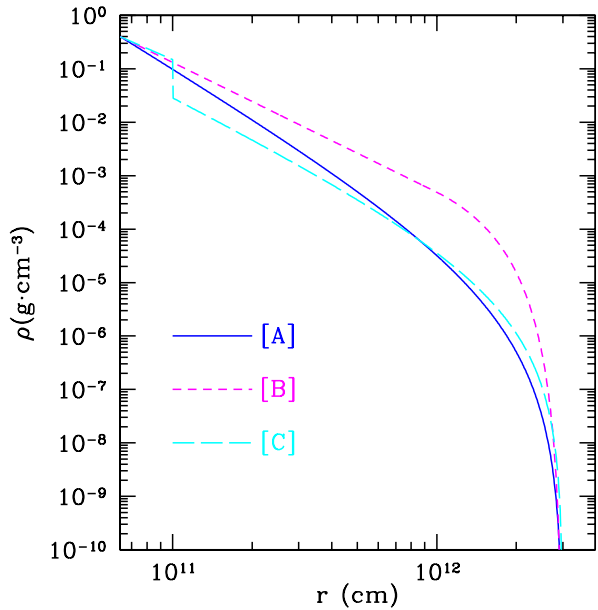}} \caption{ 
({\em Color online}) Models of outer density profile of a blue
supergiant star of radius $R_\star = 10^{12.5}$~cm. The outermost
hydrogen envelope sits on a helium envelope extending up to a radius
of $\sim 10^{11}$~cm. High-energy neutrinos are produced at a radius
$r_{\rm jet} = 10^{10.8}$~cm by a relativistic jet. The models [A],
[B] and [C] are described in Eqs. (\ref{dens-pro-A}),
(\ref{dens-pro-B}) and (\ref{dens-pro-C}) respectively.}
\label{fig:profile}
\end{figure}

Model [A] corresponds to a polytropic hydrogen envelope with $\rho (r)
\propto r^{-3}$, scaling valid in the range $r_{\rm jet} \lesssim r 
\lesssim R_\star$. Model [B] is a power-law fit with an effective 
polytropic index $n_{\rm eff} = 17/7$ as done for SN 1987A in
Ref. \cite{chevsok89}. Model [C] includes a sharp drop in density at
the edge of the helium core. Note that our choice of different $n_{\rm
eff}$ below and above the helium core, in this case, is motivated by
Ref. \cite{matzmckee99}. The parameter ${\cal A}$ corresponds to the
drop in density and its value is set by hand.
\begin{widetext}
\ba
{\rm [A]} ~~\rho(r) &=& 4.0\cdot 10^{-6} \left( \frac{R_\star}{r} -1
\right)^3 ~{\rm g~cm}^{-3} \label{dens-pro-A} \\
{\rm [B]} ~~\rho(r) &=& 3.4\cdot 10^{-5} \times \begin{cases} 
(R_\star/r)^{17/7} ~;~10^{10.8} ~{\rm cm}<r<r_b = 10^{12} ~{\rm cm} \cr
(R_\star/r_b)^{17/7} (r-R_\star)^5/(r_b - R_\star)^5 ~;~ r>r_b
\end{cases} ~{\rm g~cm}^{-3} \label{dens-pro-B} \\
{\rm [C]} ~~\rho(r) &=& 6.3\cdot 10^{-6} {\cal A}
\left( \frac{R_\star}{r} -1 \right)^{n_{\rm eff}}  
~{\rm g~cm}^{-3} ~;~ (n_{\rm eff}, {\cal A})= \begin{cases} 
(2.1,20) ~;~ 10^{10.8} ~{\rm cm}< r < 10^{11} ~{\rm cm} \cr
(2.5,1) ~;~ r > 10^{11} ~{\rm cm} \end{cases} 
\label{dens-pro-C}
\ea
\end{widetext}
We have plotted the density profiles in Eqs. (\ref{dens-pro-A}),
(\ref{dens-pro-B}) and (\ref{dens-pro-C}) in
Fig. \ref{fig:profile}. The number density of electrons is given by
$N_e(r) = N_A\rho(r)Y_e$, where $Y_e \approx 1$ in our cases, is the
number of electrons per nucleon or the electron fraction. As for
reference, $N_e = 2.4\cdot 10^{23}$ cm$^{-3}$ at $r=r_{\rm jet} =
10^{10.8}$~cm for all models described above.

We use the pion and kaon decay neutrino flux models from hadronic
($pp$) interactions by shock accelerated protons in the
semi-relativistic hidden jets \cite{Razzaque:2004yv,Ando:2005xi}. The
fluxes ($\Phi_{\nu_\mu} = \Phi_{\bar \nu_\mu} = 2 \Phi_{\nu_e}=2
\Phi_{\bar\nu_e}$), from an isolated source, at the
production site $r=r_{\rm jet}$ may be parametrized as \cite{rmw05}
\ba
\Phi_{\pi, \nu_{\mu}}^{\rm s} &=& 4\cdot 10^{45} 
\left( \frac{d_L}{\rm cm} \right)^{-2} 
\left( \frac{E_\nu}{300~{\rm GeV}} \right)^{-4} \nonumber \\
&& ~{\rm GeV}^{-1} {\rm cm}^{-2} {\rm s}^{-1} ~;~ \nonumber \\
&& ~100 \lesssim E_\nu/{\rm GeV} \lesssim 3\cdot 10^5,
\label{pi-decay-flux} \\
\Phi_{K, \nu_{\mu}}^{\rm s} &=& 8\cdot 10^{38} 
\left( \frac{d_L}{\rm cm} \right)^{-2} 
\left( \frac{E_\nu}{5\cdot 10^4~{\rm GeV}} \right)^{-\kappa} \nonumber \\
&& ~{\rm GeV}^{-1} {\rm cm}^{-2} {\rm s}^{-1} ~;~ \nonumber \\
&& ~\kappa =\begin{cases} 3 ~;~
100 \lesssim E_\nu/{\rm GeV} \lesssim 5\cdot 10^4 \cr 
4 ~;~ 5\cdot 10^4 \lesssim E_\nu/{\rm GeV} \lesssim 3\cdot 10^5,
\label{kaon-decay-flux}
\end{cases}
\ea
respectively from pion and kaon decays. Here $d_L$ is the luminosity
distance of the source. Next we discuss oscillation effects on these
neutrinos as they propagate from inside the stellar interior to the
surface in dense media, from the source to Earth and through the
Earth. Note that high energy neutrinos are produced in shocked
material in the jet which has a density lower than the surrounding
stellar material. However, the width of the shock is too small (jet
radius divided by the Lorentz boost factor of the jet) to have any
significant oscillation effect before the neutrinos start moving
through the jet head (which is roughly in equilibrium with surrounding
material) and the envelope. Also, the jet is well outside the stellar
core which may be turbulent and the envelope is not disrupted yet.

\section{Neutrino oscillation in vacuum and in matter}
\label{sec.oscillation}

Solar, atmospheric, accelerator and reactor neutrinos have provided
ample evidence for neutrino oscillations.  Neutrinos from
astrophysical sources like those described above are affected by
oscillations, which change the flavor composition of the fluxes
between production and detection.

The neutrino flavor eigenstates $\nu_\alpha$, where
$\alpha=e,\mu,\tau$ in the case of 3 flavor mixing, are related to the
mass eigenstates $\nu_j$, where $j=1,2,3$ corresponding to the masses
$m_j$, by the Maki-Nakagawa-Sakata (MNS) unitary mixing matrix $U$ as
$\nu_\alpha = U_{\alpha j} \nu_j$. The sum over repeated indices is
implied. We use the standard expression of $U$ from Ref.~\cite{pdg04}.

We use standard oscillation parameters obtained from global fits to
neutrino oscillations data:
\ba
\Delta m^2_{\rm sol}&=&\Delta m^2_{21}\sim 8\times 10^{-5}\ev^2\nonumber
\\
\theta_{\rm sol}&=&
\theta_{12}\sim 33.83^\circ\nonumber\\
|\Delta m^2_{\rm atm}|&=&|\Delta m^2_{32}|\sim 2.4\times
10^{-3}\ev^2\nonumber\\
\theta_{\rm atm}&=&\theta_{23}\sim 45^\circ\nonumber\\
\sin^2(2\theta_{13})&\aprle& 0.15
\ea
The sign of the atmospheric mass difference has not been
determined. Positive (negative) $\Delta m^2_{\rm atm}$ correspond to
the normal (inverted) hierarchies. 

Uncertainties in the oscillation parameters are still relatively large
and small variations due to these uncertainties on the generic results
we present here are to be expected. We will explore the impact of the
expected errors of the oscillation parameters on our observable and we
will discuss how these errors could affect an eventual measurement of
the neutrino mass hierarchy (i.e normal vs inverted) via matter
effects.
 
We will also consider the effects of a CP violating phase which is at
present unconstrained. A precise approximation (for constant density
and energies high enough such that $\Delta_{21}/V_e\ll 1$ and
$\Delta_{21} L\ll 1$) for the $\nu_e \rightarrow \nu_\mu$ transition
probability is given by~\cite{golden}:
\begin{eqnarray}
& P_{\nu_ e \nu_\mu ( \bar \nu_e \bar \nu_\mu ) } = \nonumber \\ &
\sin^2\theta_{23}\sin^2 2 \theta_{13} \left ( \frac{ \Delta_{31} }{
\tilde V_\mp } \right )^2 \sin^2 \left( \frac{ \tilde V_\mp L}{2}
\right) \nonumber \\ & + \cos^2 \theta_{23} \sin^2 2 \theta_{12}
\left( \frac{ \Delta_{21} }{V_{e}} \right )^2 \sin^2 \left(
\frac{V_{e}L}{2} \right)
\label{eq:complete}\\
& + \tilde J \frac{ \Delta_{21} }{V_{e}} \frac{ \Delta_{31} }{ \tilde V_\mp } 
  \sin \left( \frac{ V_{e} L}{2}\right) 
  \sin \left( \frac{\tilde V_{\mp} L}{2}\right) 
   \cos \left( \pm \delta - \frac{ \Delta_{31} L}{2} \right )~, 
\nonumber
\end{eqnarray}
where $\Delta_{ij} = \Delta m^2_{ij}/2E_\nu$, $\tilde V_\mp \equiv
|V_{e} \mp \Delta_{31}|$, the matter potential $V_{e}=V_e(r)$, $\tilde
J$ is the Jarslog invariant and the sign minus (plus) refers to
neutrinos (antineutrinos).
 
For the above values of $\Delta m^2$, even for the high energies of
interest here, the distances relevant for the sources described above
are so large that, after propagation through vacuum, only the averaged
oscillation is observable for the simple two flavor case { with the
probability given by}
\beq
P(\nu_\alpha\to\nu_\beta)=
\sin^2(2\theta)\sin^2\left(\frac{\Delta m^2
L}{4E}\right)
\longrightarrow\!\!\!\!\!\!\!\!\!\!\!\!{}_{{}_{_{L \to\infty}}}
\frac{1}{2}\sin^2(2\theta)\, .
\enq
In the three flavor case, given fluxes at the surface of the source
$\phi^0_{\nu_e}$, $\phi^0_{\nu_\mu}$ and $\phi^0_{\nu_\tau}$, the
fluxes at Earth are given by:
\ba{}\!\!\!
\phi_{\nu_e}\!\!\!&=&\!\!\!\phi^0_{\nu_e}-\frac{1}{4}\sin^22\theta_{12}
(2 \phi^0_{\nu_e}-
\phi^0_{\nu_\mu}-
\phi^0_{\nu_\tau})
\\
\!\!\!\phi_{\nu_\mu}\!\!\!&=&\!\!\!\phi_{\nu_\tau}\!\!=\!\!\frac{1}{2}(\phi^0_{\nu_\mu}+\phi^0_{\nu_\tau})+\frac{1}{8}\sin^22\theta_{12}
(2 \phi^0_{\nu_e}-\phi^0_{\nu_\mu}-\phi^0_{\nu_\tau})\nonumber
\label{vacav}
\ea
if we assume that $\theta_{23}$ is maximal and $\theta_{13}$ is very
small (as indicated by neutrino oscillation data). If neutrinos are
produced in pion {and/or kaon} decays, the initial flavor ratio is
given by $\nu_e:\nu_\mu:\nu_\tau= 1:2:0$. After propagation over very
long distances in vacuum, neutrino oscillations change this ratio to
$1:1:1$ because of the maximal $\nu_\mu\leftrightarrow\nu_\tau$
mixing.  For the sources we are considering, in the low energy range,
even the distance traveled inside the source is large enough that
neutrinos oscillate many times and the phase information is lost.
When the initial ratio $\nu_e:\nu_\mu:\nu_\tau$ is different from
$1:2:0$, the flavor ratio at Earth is affected by the full three
flavor mixing and is different from $1:1:1$.

In the case of interest here, even though neutrinos are produced by
pion and kaon decays, the fluxes at Earth can have a flavor
composition quite different from the standard $1:1:1$ ratio. This is
because the neutrino flavor composition is affected by the propagation
through matter inside the astrophysical object.

The fluxes at the surface of the star are given by {the sum of
products of the fluxes at the production site
[Eqs.(\ref{pi-decay-flux}) and (\ref{kaon-decay-flux})] and the
oscillation probabilities as}
\ba
\phi^0_{\nu_e}&=&\phi^{\rm s}_{\nu_e}P(\nu_e\to\nu_e) + \phi^{\rm s}_{\nu_\mu}
P(\nu_\mu\to\nu_e) \nonumber\\ &=&\phi^{\rm s}_{\nu_\mu}
\left(\frac{1}{2} P(\nu_e\to\nu_e)+
P(\nu_\mu\to\nu_e)\right)\nonumber\\
\phi^0_{\nu_\mu}&=&\phi^{\rm s}_{\nu_\mu} \left( \frac{1}{2}P(\nu_e\to\nu_\mu)+ P(\nu_\mu\to\nu_\mu)\right)\nonumber\\
\phi^0_{\nu_\tau}&=&\phi^s_{\nu_\mu} \left(\frac{1}{2}P(\nu_e\to\nu_\tau)+P(\nu_\mu\to\nu_\tau)\right)
\ea
The same type of relations apply to anti-neutrinos.

The $\nu_\alpha \leftrightarrow \nu_\beta$ oscillation probabilities
$P_{\alpha\beta}$ on the stellar surface $r=R_\star$ can be obtained
by solving, with the appropriate initial conditions, the evolution
equation:
\begin{widetext}
\ba
i\frac{d}{dr} \left( \begin{array}{c} \nu_e \\ \nu_{\mu} \\
\nu_{\tau} \end{array} \right) = \left[ \frac{1}{2E} U 
\left( \begin{array}{ccc} 0 & 0 & 0 \\
0 & \Delta m_{21}^2 & 0 \\ 
0 & 0 & \Delta m_{31}^2 \end{array} \right) U^\dagger +
\left( \begin{array}{ccc} V_e(r) & 0 & 0 \\ 0 & 0 & 0 \\ 0 & 0 & 0 \end{array} 
\right) \right] \left( \begin{array}{c} \nu_e \\ \nu_{\mu} \\ \nu_{\tau}
\end{array} \right).
\label{evol-eq}
\ea
\end{widetext}
Here $V_e(r)=\sqrt{2} G_F N_e (r)$ is the matter-induced charged
current potential for $\nu_e$ in a medium of electron number density
$N_e (r)$ found from Eqs. (\ref{dens-pro-A}), (\ref{dens-pro-B}) and
(\ref{dens-pro-C}). For $\bar \nu_e$ the potential changes sign.

Matter effects are expected to be large when neutrinos go through a
resonant density. For constant electron density, there would be two
such resonances, corresponding to the two mass squared differences:
\ba
N_e^L&=&\frac{\Delta m^2_{\rm sol}\cos2\theta_{12}}{2\sqrt{2}G_F
E}\simeq\frac{3\times 10^{23}}{E[GeV]}{\rm cm}^{-3}\label{sol}\\
N_e^H&=&\frac{|\Delta m^2_{\rm atm}|\cos2\theta_{13}}{2\sqrt{2}G_F E}
\simeq\frac{9\times 10^{24}}{E[GeV]}{\rm cm}^{-3}\label{atm}
\ea
As can be seen from the previous section, these densities can be
encountered in the sources of interest by neutrinos with the relevant
energies.

Even if $\theta_{13}$ is very small, matter effects can be expected
from the solar $\Delta m^2$ oscillations [see Eq. (\ref{sol})].  If
$\theta_{13}$ is non-negligible, larger matter effects can be expected
from the atmospheric $\Delta m^2$ oscillations [see
Eq. (\ref{atm})]. These are non-negligible at a relatively higher
energy.

Since the matter potential is different for anti-neutrinos, it is
important to independently solve the evolution equations for neutrinos
and anti-neutrinos and consider the averaging over the two fluxes only
when reaching the detector, which cannot distinguish between the
two. Note that if the hierarchy is normal (inverted) the resonant
behavior occurs only in the neutrino (anti-neutrino) channel.

The matter density in the sources we are considering is varying rather
strongly, such that a resonance is not directly observable.  Adiabatic
conditions are not usually satisfied, such that a full numerical
treatment of the problem is necessary. The adiabatic approximation can
still be used in some energy range. It is useful to understand some of
the features introduced by matter effects by analyzing limiting
regions of the models described in the previous section.  The
adiabaticity parameter is defined as (see Ref.\ \cite{Dighe:1999bi}
and references therein):
\beq
\gamma \equiv \frac{\Delta m^2}{2 E} 
\frac{\sin^2 2\theta}{\cos 2\theta}
\frac{1}{(1/N_e)(dN_e/dr)}~,
\enq
i.e.\ the ratio of the resonance width and the neutrino oscillation
length. In other words, the adiabaticity parameter represents the
number of oscillations that occur in the resonance region. Adiabatic
transition ($\gamma \gg 1$) means that the neutrino will remain in its
particular superposition of initial instantaneous mass eigenstates as
it crosses the resonance. The ``flip'' probability, that is, the
probability that a neutrino in one matter eigenstate jumps to the
other matter eigenstate is
\beq
P_{\rm flip} = \exp \left(-\frac{\pi}{2} \gamma \right)~,
\enq
which means that, if the neutrino adiabaticity parameter is $\gamma\gg
1$, the ``flip'' probability $P_{\rm flip}\simeq 0$, many oscillations
will take place at the resonance region and strong flavor transition
will occur.  On the other hand, if the adiabaticity parameter is
$\gamma \geq 1$, only a few oscillations take place in the resonant
region and the mixing between instantaneous matter eigenstates is
important, and $P_{\rm flip}$ gives the jumping probability from one
instantaneous mass eigenstate to another.  For the solar transition,
where the mixing angle is large, the appropriate expressions for the
adiabaticity conditions and flip probabilities can be found in
\cite{flipsolar}. Since the main effects that we are interested in
correspond to the small mixing angle, we limit our analitical
discussion to this case. We would like to emphasize that in order to
obtain the correct neutrino oscillation probabilities a full numerical
treatment of the problem is necessary and our discussion based on flip
probability serves only as a qualitative guide to understanding the
solution in some limited energy ranges where adiabaticity is
satisfied.

\begin{figure} [ht]
\centerline{\includegraphics[width=3.4in]{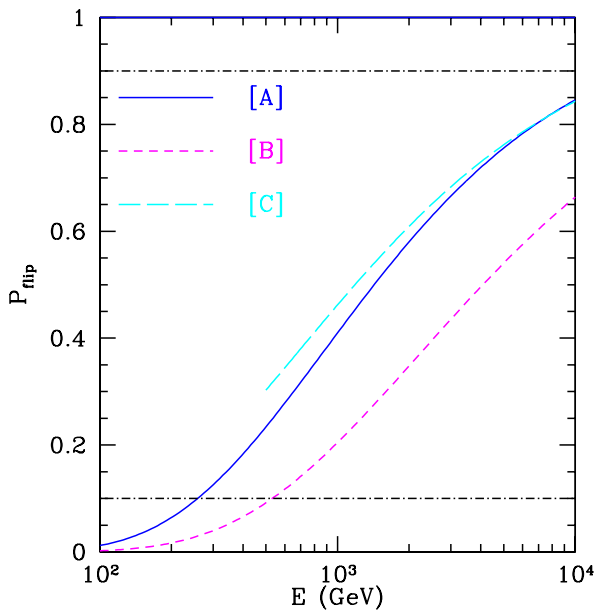}} 
\caption{\label{fig:flip} ({\em Color online}) 
Flip probability corresponding to the atmospheric transition for
$\sin^22\theta_{13}=0.15$, for the three density profiles in
Fig.\ \ref{fig:profile}.}
\end{figure}

Figure \ref{fig:flip} shows the flip probability corresponding to the
atmospheric transition for the three density profiles we consider.
We explicitly delimit three different regions: the adiabatic region
corresponds to very small flip probabilities ($P_{\rm flip} <0.1$) and
strong flavor transitions; in the intermediate region ($0.1<P_{\rm
flip}<0.9$) the transitions are not complete; the highly non-adiabatic
region corresponds to very high flip probabilities ($P_{\rm
flip}>0.9$).

At low energies transitions are adiabatic for the density profiles in
models [A] and [B], so strong conversion is expected for these two
models. Model [C] has a sharp drop in density and at energies below
500 GeV the resonant density is reached only on the step, where it is
highly non-adiabatic, so one does not expect a significant matter
enhancement.

At higher energies the adiabatic regime applies only for Model [B],
for which oscillation effects are expected to be large, while for
models [A] and [C] the flavor transition is incomplete and a smaller
effect is expected. These expectations are confirmed by our exact
results, that is, around energies $\sim 1$~TeV, the effect for the
Model [A] and the Model [C] should be similar and smaller than the one
observed for the Model [B].  The differences in the adiabaticity
behavior for the three models are induced by their different matter
profiles, see Fig.\ \ref{fig:profile}: for Model [B] (which has the
least steep matter density), the adiabaticity condition is satisfied
for a larger neutrino energy range. The adiabaticity of the
transitions depends on the matter density profile. If this is very
steep, the corresponding resonance width will be very small (it is
inversely proportional to the derivative of the matter potential) and
consequently the neutrino system can not adapt itself, only a few
oscillations will take place and the flavor transitions will not be
complete. In order to have large matter effects it is also necessary
to go through a minimum matter width, as shown in Ref.\
\cite{lunardinismirnov}. For the sources we consider here the density
is high enough that this condition is satisfied.

\begin{figure} [ht]
\centerline{\includegraphics[width=3.5in]{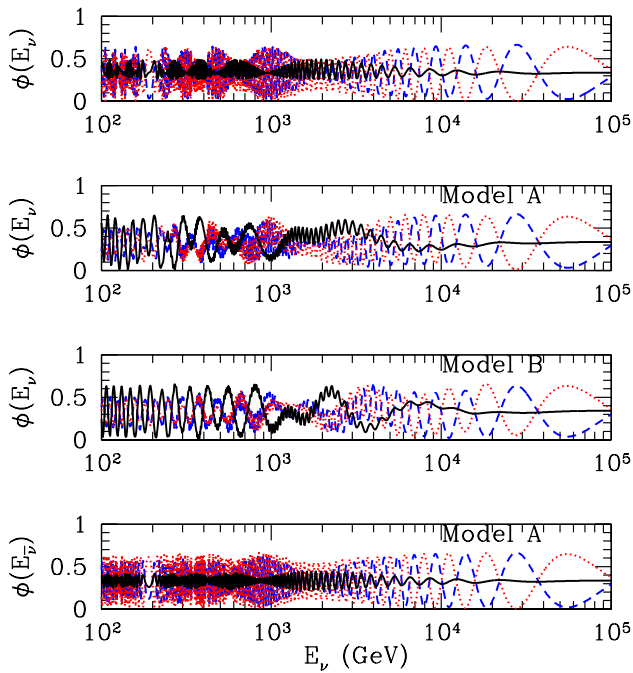}}
\caption{ ({\em Color online})
Neutrino fluxes at the surface of the source (for
$\sin^22\theta_{13}=0.15$), normalized to the initial electron
neutrino fluxes. From upper to lower panels: vacuum, Model [A], Model
[B] and Model [A] antineutrinos. The solid, dashed and dotted curves
are for $\nu_e$, $\nu_\mu$ and $\nu_\tau$ respectively. }
\label{fig:surface}
\end{figure}

In Fig.\ \ref{fig:surface} we show the electron, muon and tau neutrino
fluxes at the surface of the source, normalized to the initial
electron neutrino fluxes.  We compare the case when only the vacuum
oscillations are considered inside the source with the fluxes of
neutrinos when matter effects are taken into account with a density
profile as in Model [A] or Model [B], for normal hierarchy. The
averaging due to fast vacuum oscillations can already be observed in
the lower energy range. It can also clearly be seen that matter
effects modify the flavor composition of the neutrino fluxes,
introducing energy-dependent features. As expected from the discussion
above, the effects are larger for Model [B] where adiabatic-like
transitions occur in a wider energy range. We also show the
corresponding anti-neutrino fluxes for Model [A] {for comparison}. The
results are very close to those in vacuum, as expected for the normal
hierarchy.

After propagating the neutrinos through matter to the surface of the
star, the fluxes at Earth immediately follow by considering the
averaged oscillations over the long travel distance in vacuum as
obtained from Eq. (\ref{vacav}).

\begin{figure} [ht]
\centerline{\includegraphics[width=3.5in]{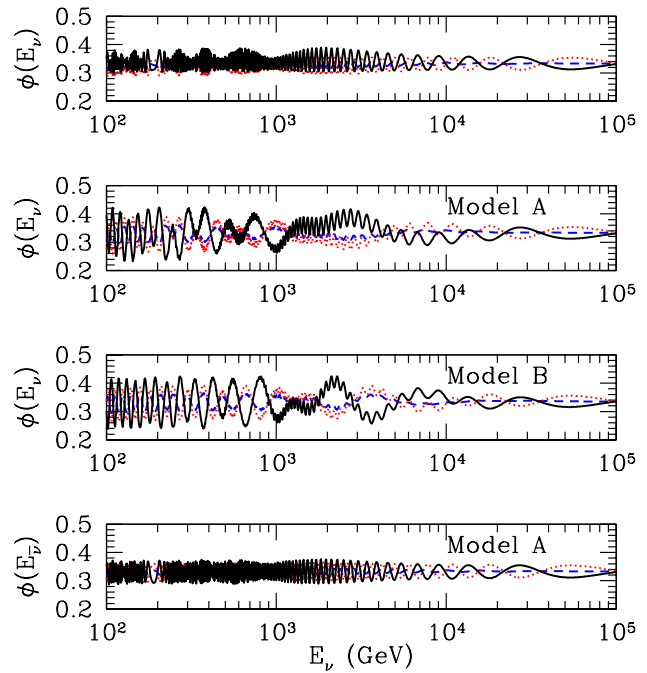}}
\caption{ ({\em Color online})
Same as Fig.\ \ref{fig:surface}, but at the surface of the Earth.}
\label{fig:Earth}
\end{figure}

Figure \ref{fig:Earth} makes the same comparison presented in Fig.\
\ref{fig:surface} after propagation all the way to the Earth. The
flavor composition of the neutrinos gets further modified by the
propagation from the source to the Earth.

\section{Detection}
\label{sec.detection}

On their way to detectors, neutrinos also propagate through matter in
the Earth. In this case matter effects on neutrino oscillations are
however extremely small, since the energies considered here are much
higher than the resonant energy inside the Earth. For $\Delta m^2_{\rm
atm}=2.4 \times 10^{-3}$ eV$^2$ the corresponding resonant energy for
a neutrino going through the Earth's mantle (with the density $\sim 3$
g/cm$^3$) is around 10 GeV, already lower than the energies we
consider. The resonant energy is even lower for the solar mass
difference or for trajectories going through the higher density Earth
core \cite{MS}.

The charged current and neutral current neutrino-nucleon interaction
cross-section increases with energy and attenuation effects start
becoming important for the propagation through matter. The interaction
length:
\beq
{\cal L}=\frac{1}{\sigma_{\nu N} (E)  N_A}
\enq
becomes comparable to the Earth diameter around 40 TeV. The
attenuation effects are thus becoming relevant only at the highest
energies considered here, where the fluxes are very
small. Consequently, propagation through the Earth will have
negligible effects on the neutrino fluxes and flavor composition
previously discussed.

Neutrinos from the sources discussed here are expected to be
detectable in IceCube and other neutrino telescopes which have good
sensitivity in the energy range of interest, between 100 GeV and 100
TeV.

It is possible in IceCube to get information about the flavor
composition of the neutrinos as well. The detector is mostly sensitive
to observing muons identified by their very long tracks, thus counting
the number of $\nu_\mu$ interactions. Electron neutrino interactions
create electromagnetic cascades which can be observed in IceCube.  Tau
neutrinos will also lead to cascades. At these energies the tau decay
length is very small and the interaction of the neutrino and tau decay
cannot be separated. It might be possible to separately infer the
fluxes of all three flavors if hadronic showers can be well separated
from electromagnetic showers through their muon track content. We will
consider taus to be indistinguishable from electrons in our analysis
and we will compare the number of tracks due to $\nu_\mu$ and number
of showers due to $\nu_e$ and $\nu_\tau$.

Another important feature of interest for the detection of the effects
studied in this paper is the energy dependence of the signal: as
discussed in the previous section, matter effects have a strong
dependence on energy, which is correlated with the density profile
inside the source; changes in neutrino oscillation parameters also
induce energy dependent effects, as we discuss in more detail in the
next section. The energy resolution in IceCube is expected to be
somewhat better than 30\% on a logarithmic scale for muon neutrinos
and about 20\% on a linear scale for cascades.

We explore the spectral shape of the shower-to-muon track ratio
inferred at Icecube, defined as:
\beq
R=\frac{N_{\nu_e} + N_{\bar{\nu}_e} +N_{\nu_\tau} +
N_{\bar{\nu}_\tau}}{N_{\nu_\mu} + N_{\bar{\nu}_\mu}}~.
\label{eq:ratio}
\enq
The absolute flux normalization thus does not affect our results.
The number of events, $N_{\nu_{\alpha}(\bar{\nu}_{\alpha})}$ is
proportional to:
\beq
N_{\nu_{\alpha}(\bar{\nu}_{\alpha})} \sim \int_{\Delta E_\nu}
(\phi^{\pi}_{\nu_{\alpha}(\bar{\nu}_{\alpha})} +
\phi^{K}_{\nu_{\alpha}(\bar{\nu}_{\alpha})})(E_\nu) \times
\sigma_{\nu_{\alpha}(\bar{\nu}_{\alpha})} (E_\nu) dE_\nu~,
\enq
i.e.\ we have computed the (anti)neutrino fluxes from pion and kaon
decays at the detector after considering the propagation inside the
source (affected by matter effects) as well as the propagation from
the source to the detector. The (anti)neutrino fluxes are then
convoluted with the (anti)neutrino cross-sections~\cite{gqrs} and
integrated over the neutrino energy, assuming a conservative energy
bin size of $\Delta E_\nu=0.3E_\nu$. If the matter potential inside
the astrophysical source is neglected, the neutrino flavor ratio in
Eq.~(\ref{eq:ratio}) should be almost constant, $R\simeq2$, over all
the neutrino energy range. Adding the matter contribution to the
neutrino propagation will change this constant value of $R$ in an
energy dependent fashion. In the following section we explore the
spectral behavior of the shower-to-track ratio in the 300 GeV- 300 TeV
neutrino energy range, considering both normal and inverted neutrino
mass orderings and exploring the effects of a non zero leptonic CP
violating phase $\delta$ and of uncertainties in other oscillation
parameters.

\section{Spectral Analysis}
\label{sec.analysis}

We present here the main results of our study. We are mainly
interested in neutrino propagation through matter inside the
source. This is followed by the propagation through vacuum between the
source and Earth. Matter effects and absorption inside the Earth are
negligible for the energies considered here. As discussed in the
previous section, a good observable for studying the effects of
neutrino propagation inside the source is the spectral shape defined
in Eq.~(\ref{eq:ratio}). While $R\simeq 2$ for the optically thin
sources, energy dependent deviations from this value are expected due
to matter effects in the astrophysical hidden sources considered here.

\begin{figure} [ht]
\centerline{\includegraphics[width=3.4in]{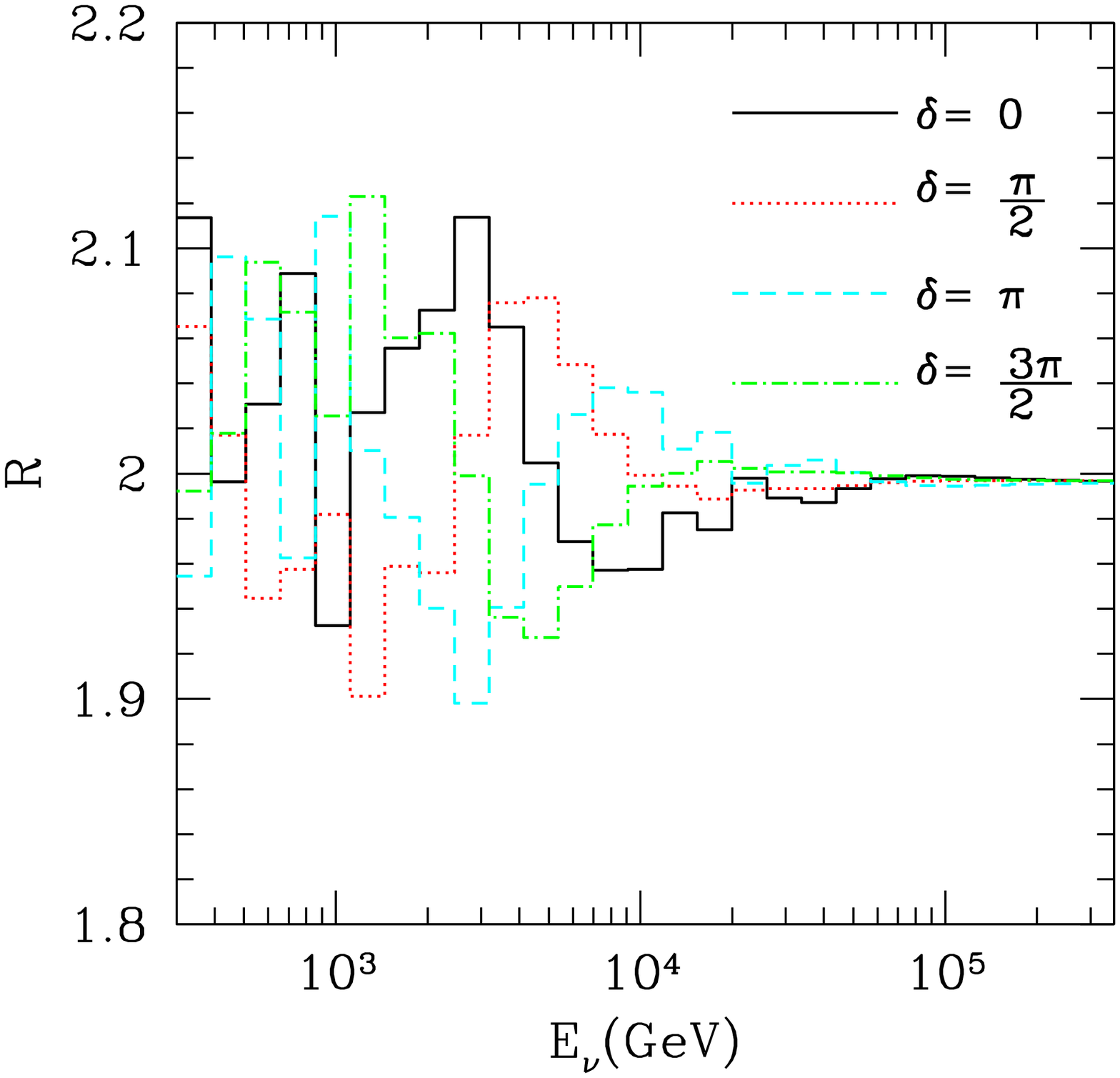}}
\centerline{\includegraphics[width=3.4in]{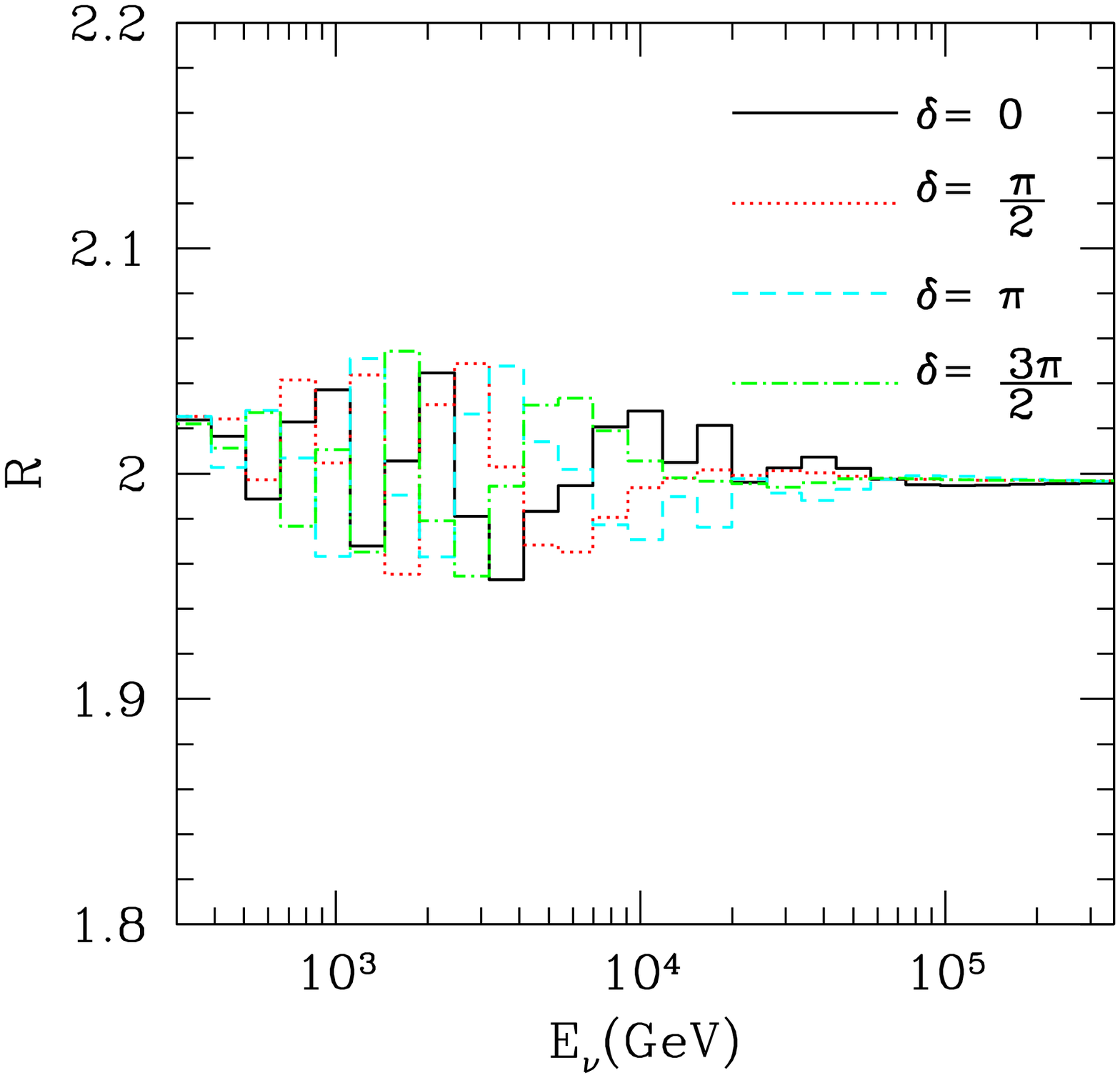}}
\caption{\label{fig:modela} ({\em Color online})
The upper (lower) panel depicts the shower-to-muon track ratio defined
in Eq. (\ref{eq:ratio}) for $\sin^22\theta_{13}=0.15$, for Model [A],
for the case of normal (inverted) hierarchy and for different values
of the CP violating phase $\delta$.}
\end{figure}

Figure~\ref{fig:modela} shows the shower-to-muon track ratio in
Eq.~(\ref{eq:ratio}) for the normal and inverted neutrino mass
hierarchies for the Model [A] density profile [see
Eq.~(\ref{dens-pro-A})]. Deviations from the standard scenario are
non-negligible in the energy range around a few TeV. The impact of the
matter potential is much larger if nature has chosen the normal mass
ordering, since, if that is the case, only neutrinos can go through
the resonance density $N^{H}_e$ and the effect will be larger due to
the higher neutrino cross-section, when compared to the antineutrino
one. For neutrino energies $E_\nu< 6 $ TeV, where the resonant effect
in the (anti)neutrino propagation is expected to be located, the
neutrino cross-section is roughly twice the antineutrino cross
section. Therefore, the matter potential impact in the inverted
hierarchy situation (when only the antineutrino propagation gets
distorted) is much smaller than in the normal hierarchy case.  We have
also studied the shower-to-muon track ratio spectral shape when the CP
violating phase $\delta$ is not zero. The different curves in
Fig.~\ref{fig:modela} correspond to different values of the CP
violating phase $\delta$.  They were obtained by numerically computing
the full oscillation probabilities, but the relative shape of these
curves with respect to the $\delta=0$ curve shape can be easily
explained by means of the oscillation probabilities in
Eq. (\ref{eq:complete}). This is because the expression applies for
almost all the relevant energies and baselines explored in our
study. It can be easily seen, for example, that for a fixed distance
of $L=10^{5}$ km and a (constant) density correspnding to the electron
number density for Model A, $\Delta_{12}/V_{e}\ll 1 $ and $\Delta_{12}
L < 1$ for energies $\sim 10^{3}$ GeV and the approximated probability
formula provides an accurate description of the oscillation
probabilities versus the CP phase $\delta$.

There are three very different terms in Eq.~(\ref{eq:complete}): the
first one, which is the dominant one, is responsible for the matter
effects; the second one, which is the
\emph{solar} term, is dominant for very small values of $\theta_{13}$
not explored here, and therefore, negligible for the present
discussion; the third term, (named the
\emph{interference} term in the literature), is the only one
which depends on the CP violating phase $\delta$ and it determines the
shape of the curves in Fig.~\ref{fig:modela}. In the case of neutrino
transitions and normal hierarchy, notice that if $\delta=0$, the three
oscillatory factors of the third term can be written as $\sin^2
(\tilde V_{-}L/2)$. If, for instance, $\delta=\pi$, the three
oscillatory functions can be written as $-\sin^2 (\tilde V_{-}L/2)$,
which explains why when the $\delta=0$ curve has a maximum and the
$\delta=\pi$ curve has a minimum. If $\delta=\pi/2$, the three
oscillatory factors can be simplified as $\sin \tilde (V_{-}L/2) \cos
(\tilde V_{-} L/2)$, which explains the relative shift in the phase of
the $\delta=\pi/2$ curve with respect to the $\delta=0, \pi$
curves. Finally, if $\delta = 3\pi/2$, the oscillatory dependence goes
as $-\sin (\tilde V_{-}L/2) \cos (\tilde V_{-} L/2)$, i.e, it is the
opposite to $\delta=\pi/2$.  In summary, one can predict the shape
dependence on the CP phase of the shower-to-muon track ratio just by
observing the difference in the oscillation pattern of the four
functions $\pm \sin^2x,\pm\sin(x)\cos(x)$.

In the case of the inverted hierarchy, see Fig.~\ref{fig:modela}
(lower panel), the situation is reversed: notice, [see
Eq.~(\ref{eq:complete})], that a change in the sign of $\Delta
m^2_{31}$ can be traded in the vacuum limit ($V_{e}\rightarrow 0$) for
the substitution $\delta\rightarrow
\pi-\delta$. Matter effects, however, can break this degeneracy. By
making use of the matter effects, in principle, it could be possible
to extract the neutrino mass hierarchy, since the neutrino oscillation
probability is enhanced (depleted) for positive (negative) $\Delta
m^2_{31}$.

\begin{figure} [ht]
\centerline{\includegraphics[width=3.4in]{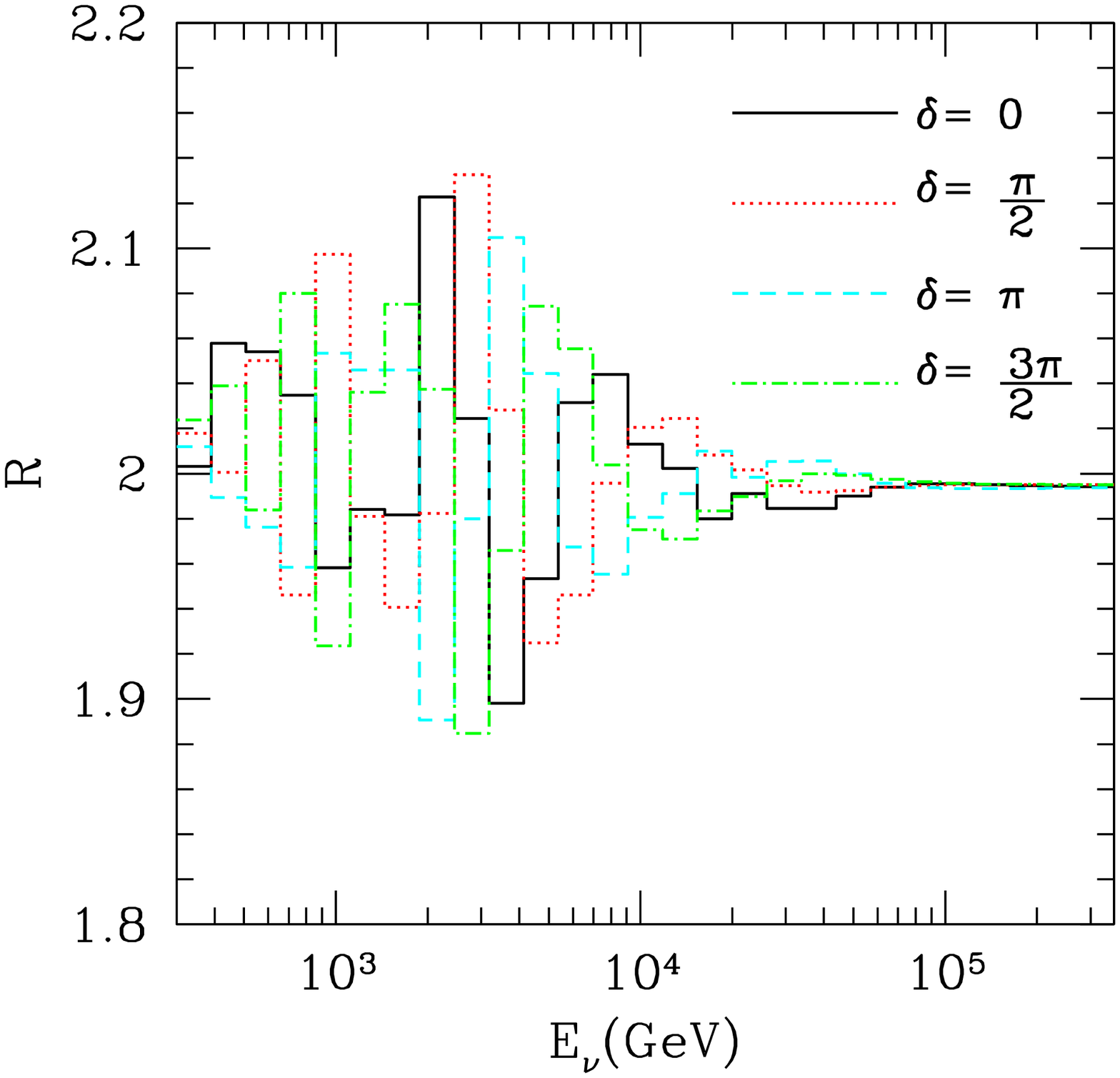}}
\centerline{\includegraphics[width=3.4in]{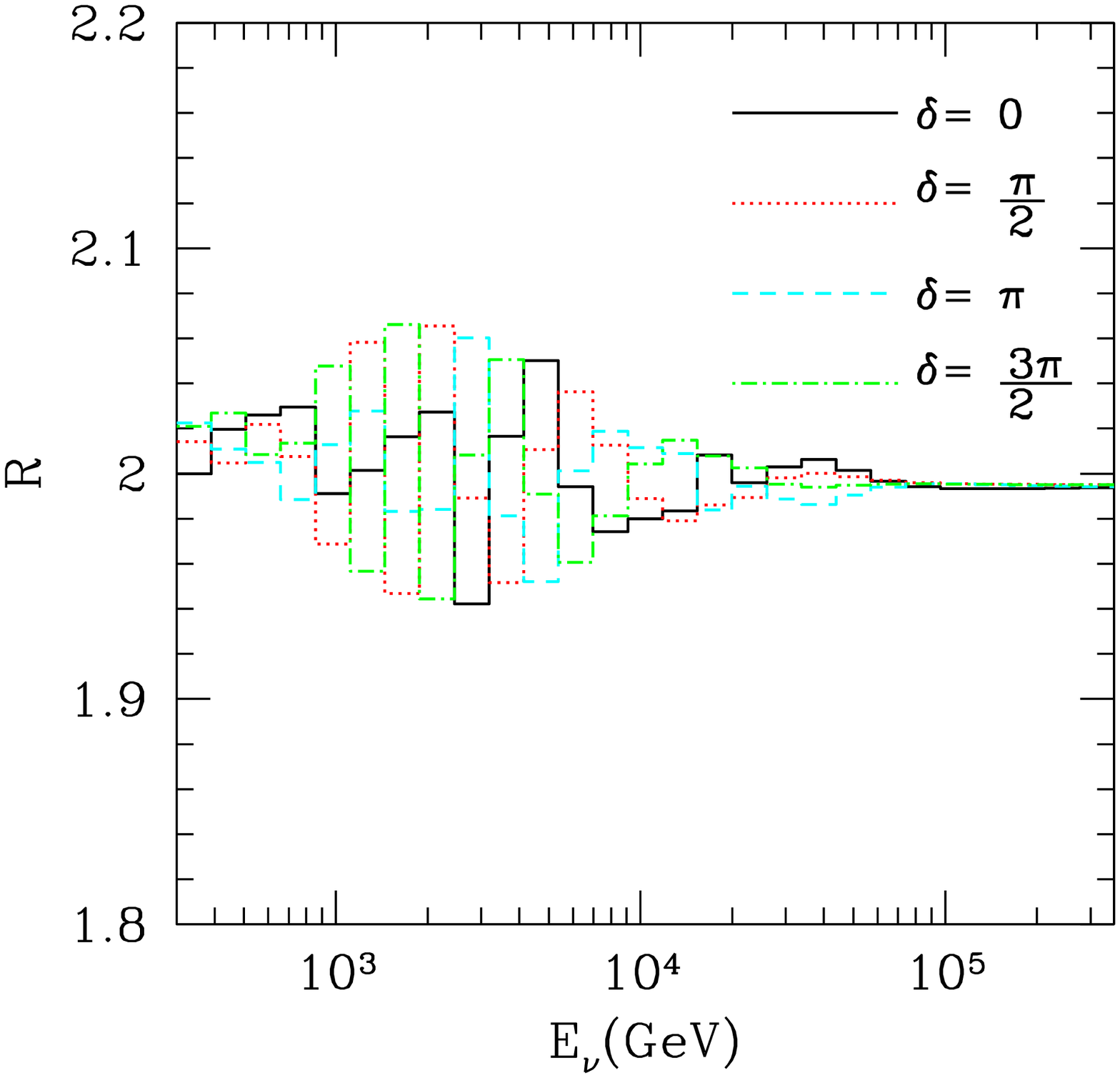}}
\caption{\label{fig:modelb} ({\em Color online})
The upper (lower) panel depicts the shower-to-muon track ratio for
$\sin^22\theta_{13}=0.15$, for Model [B], for the case of normal
(inverted) hierarchy, for different values of the CP violating phase
$\delta$.}
\end{figure}

\begin{figure} [ht]
\centerline{\includegraphics[width=3.4in]{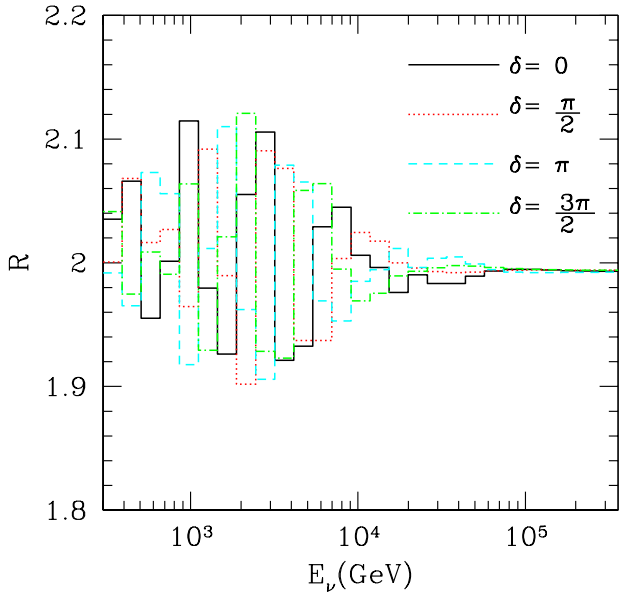}}
\centerline{\includegraphics[width=3.4in]{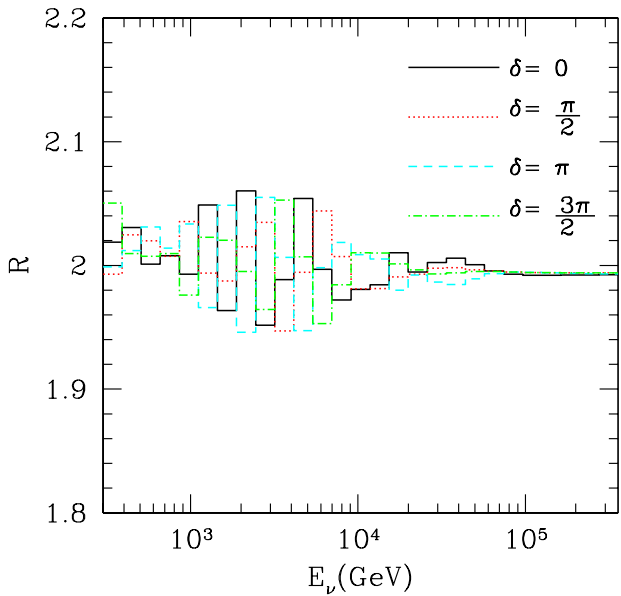}}
\caption{\label{fig:modelc} ({\em Color online}) 
The upper (lower) panel depicts the shower-to-muon track ratio for
$\sin^22\theta_{13}=0.15$, for Model [C], for the case of normal
(inverted) hierarchy, for different values of the CP violating phase
$\delta$.}
\end{figure}

Figures~\ref{fig:modelb} and \ref{fig:modelc} depict the
shower-to-muon track ratio for the density profiles given by the
models [B] and [C] in Eqs.~(\ref{dens-pro-B}) and (\ref{dens-pro-C})
respectively; and the previous discuussions on Model [A] in
Fig.~\ref{fig:modela} apply to these cases as well.

\begin{figure} [ht]
\centerline{\includegraphics[width=3.4in]{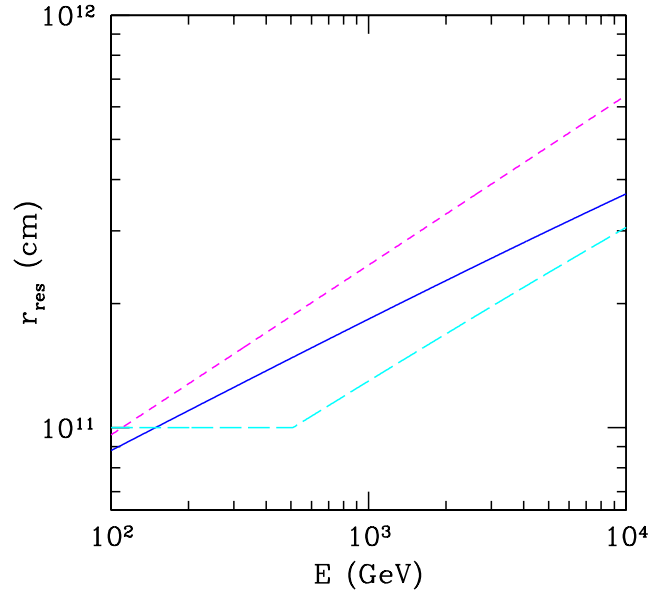}}
\caption{\label{fig:distanceenergyres} ({\em Color online})
Distance from the center of the astrophysical source (in cm), at which
the resonant density is crossed vs the neutrino energy, for Model [A]
(solid line), Model [B] (dashed line) and Model [C] (long-dashed
line), for $\sin^22\theta_{13}=0.15$.}
\end{figure}

A change in the dependence of the matter potential profile will affect
the location of the energy at which the maximum matter effect is
located. Depending on the specific form of the electron number density
vs distance $N_e(r)$, a neutrino with a fixed energy $E_\nu$ will
reach the resonant density at a different distance from the center of
the astrophysical source, as can be seen from
Fig.~\ref{fig:distanceenergyres}, where we have depicted the distance
at which the resonant density is reached vs the neutrino energy
$E_\nu$ for the three models. Notice that, for Model [C], there is a
range of energies (between 100 and 500 GeV) for which the neutrino can
profile, where the transition is highly non-adiabatic.  The
propagation inside the source from the production point to its surface
is thus 
different measured spectral shape.

It would be interesting to explore if it is possible to use neutrinos
from astrophysical sources as those discussed here in order to explore
neutrino properties like the neutrino mass ordering or the CP
violating phase $\delta$.

A deviation from the expected $1:1:1$ ratio has been extensively
explored in the literature (sometimes involving highly exotic
scenarios), e.g., a different flavor ratio at the source~\cite{rm,
Kashti:2005qa,Kachelriess:2006fi}, decaying neutrino
scenarios~\cite{bbhpw1}, Pseudo-Dirac schemes~\cite{bbhlpw},
additional sterile neutrinos~\cite{Dutta:2001sf}, magnetic moment
transitions~\cite{ekm} and the possibility of measuring deviations
from maximal atmospheric mixing~\cite{serpico}.  An extensive
discussion of possible deviations from the equal flavours scenario,
including oscillations and decays beyond the idealized mixing case,
characterization of the source and CPT violation was presented in
\cite{bb}. Some of these studies predict energy dependent ratios,
however, none discussed matter effects on high-energy neutrino flux
ratios as we explore in this paper.

\begin{figure} [ht]
\centerline{\includegraphics[width=3.4in]{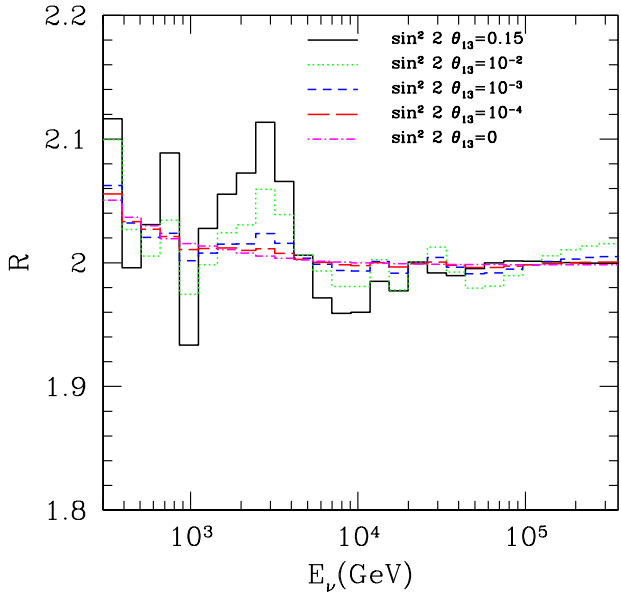}}
\caption{ ({\em Color online})
Variation of $R$ with $\theta_{13}$.
\label{fig:theta13} }
\end{figure}

If a relatively \emph{large} deviation ($\sim 10\%$) of $R$ from its
\emph{standard} value $R\simeq 2$ is observed for energies $E<2$ TeV, one
could infer a non zero value for the mixing angle $\theta_{13}$.  The
energy location of the matter enhanced \emph{peaks} in the
shower-to-muon track ratio depends on the precise value of
$\theta_{13}$: for smaller values of $\sin^22 \theta_{13}<0.15$, the
effect is larger (smaller) at lower (higher) energies with respect to
the effect illustrated here for $\sin^2 2 \theta_{13}=0.15$, as
expected from Eq.~(\ref{atm}). Figure \ref{fig:theta13} shows the
change in $R$ when varying $\theta_{13}$, for the Model A density
profile, normal hierarchy and $\delta=0$. It can be seen that the
results strongly depend on the $\theta_{13}$ mixing angle, whose
effects get enhanced by a matter resonance for values larger than
about $\sin^22 \theta_{13}=0.01$, when the transitions are mostly
adiabatic.

Measuring the CP phase by observing the shower-to-muon track ratio is
highly challenging, since for that purpose a precise knowledge of the
matter density profile $\rho(r)$ would be required (see
Ref.~\cite{bbhpw2} for the prospects in a decaying neutrino scenario,
Ref.~\cite{Winter:2006ce} for a combined analysis of neutrinos from
reactors and optically thin astrophysical sources, and
Ref.~\cite{Costantini:2004ap, Rodejohann:2006qq,
Fogli:2006jk,Xing:2006uk} for astrophysical neutrinos in different
scenarios).

\begin{figure} [ht]
\centerline{\includegraphics[width=3.4in]{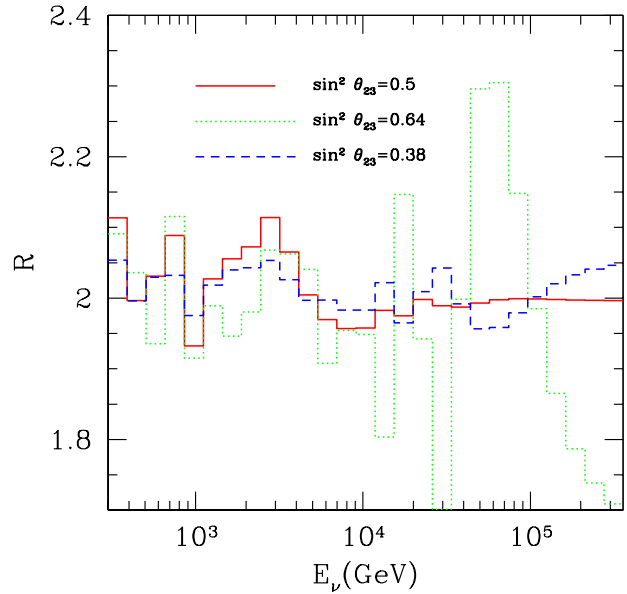}}
\caption{ ({\em Color online})
Variation of $R$ with $\theta_{23}$.
\label{fig:theta23} }
\end{figure}

Under the assumption of a quite good knowledge of the density profile,
it could be in principle possible to determine/verify the neutrino
mass hierarchy since in all three models the effect is much higher in
the normal mass ordering than in the inverted one regardless the value
of the CP violating phase $\delta$. However, in practice, the errors
on the remaining oscillation parameters might compromise the
possibility of extracting the neutrino mass hierarchy, in particular
the uncertainties on the atmospheric mixing angle $\theta_{23}$ (if a
more precise knowledge of this mixing angle is not available at the
time of the shower-to-muon track ratio measurement). Figure
\ref{fig:theta23} shows the effects of changing $\theta_{23}$ between
maximal mixing and $\sin^2 \theta_{23}=0.38$ (the current 2 sigma
allowed region), for the Model [A] density profile, normal hierarchy
and $\delta=0$. Matter effects are, in general, larger for larger
values of $\sin^2 \theta_{23}$, except for very high energies when the
ratios approach their values in vacuum. In this range, if
$\theta_{23}$ is larger than $\pi/4$ the muon flux is enhanced and
$R<2$, while the opposite is true if $\theta_{23}$ is smaller than
$\pi/4$. The uncertainty in the mixing parameters would affect the
possible determination of the mass hierarchy since the inverted
hierarchy pattern could be confused with a smaller value of
$\theta_{23}$.  A separate measurement of electron and tau neutrinos
would greatly improve this situation, allowing to extract both the
hierarchy and the value of $\theta_{23}$.

\section{Conclusions and outlook}
\label{sec.conclusion}

We have discussed here high-energy neutrinos produced in optically
thick (hidden) astrophysical objects for which large matter density
inside the source can affect the oscillations of these neutrinos. The
matter-induced transitions can lead to significant deviations from the
1:1:1 flavor ratios expected in standard scenarios. These deviations
are expected at specific energies determined by the density profiles
inside the sources. IceCube would be in an ideal position to measure
such effects. The main observable that would be sensitive to these
effects and that we have studied in detail is the shower-to-muon track
ratio defined in Eq.~(\ref{eq:ratio}).

A large number of events would be necessary in order to measure the
effects of neutrino oscillations inside the source. In order to
establish a 3 sigma effect, more than 1000 events would be needed.
This would require a relatively nearby source, within a few
megaparsecs.  Given the supernova rate in nearby starburst galaxies
such as M82 (3.2 Mpc) and NGC253 (2.5 Mpc) is about 0.1/yr, the
possibility of such an event to take place is not rare assuming a
significant fraction of SNe are endowed with jets
~\cite{Razzaque:2004yv, Ando:2005xi, rmw05}. Upcoming neutrino
telescopes will be able to constrain this fraction. The combined SN
rate from all galaxies within 20 Mpc is more than 1/yr
~\cite{Ando:2005ka}.

With such large number of events it would also be possible, in
principle, to investigate neutrino properties like the mass ordering
and CP violating phase, especially if future reactor and accelerator
experiments could provide a better measurement of the mixing
angles. If all neutrino properties would already be known from other
experiments, the shower-to-muon track ratio measured in IceCube could
be used to determine source properties: matter effects inside the
source reflected in the ratio $R$ are extremely sensitive to the
source density profile.

Since high-energy $\gamma$-rays are not emitted from optically thick
sources, it is hard to predict their occurrence rate. If a large
population of hidden high-energy neutrino sources would exist in
nature, then their combined effect would be evident in the ratio $R$
measured from diffuse fluxes. A modulation of $R$ with energy over a
wide range is expected if the astrophysics varies from source to
source and/or if the sources are distributed over a wide
redshift. This is a distinct signal from the effect, e.g., discussed
in Refs.~\cite{rm, Kashti:2005qa}; which appear only at
ultrahigh-energies.

Neutrino telescopes are sensitive to the atmospheric neutrinos as well
in the 0.1-100 TeV energy range we are interested in. However, the
atmospheric neutrinos should not affect measurement of the ratio $R$
for a nearby transient point source that we are considering. For
diffuse flux from all such point sources, the measured ratio will be a
convolution of the atmospheric flux ratios and the astrophysical flux
ratios. The oscillation effects for the atmospheric neutrinos is
rather small ($< 10\%$) because of a small $\Delta m^2 L/E$. The
experiments should be able to de-convolute them even without precise
knowledge of the oscillation parameters. Assuming neutrino oscillation
parameters are well-measured, matter oscillation effects as we
discussed here would be the most natural explanation of any deviation
of $R$ from its value in vacuum measured by upcoming neutrino
telescopes.

\section*{Acknowledgments} 

We thank Peter M\'esz\'aros for helpful discussions. Work supported by
NSF grants PHY-0555368, AST 0307376 and in part by the European
Programme ``The Quest for Unification.''  contract
MRTN-CT-2004-503369.

\end{document}